\documentclass[aps,reprint,showpacs,superscriptaddress,
footinbib,citeautoscript,floatfix]{revtex4-1}

\usepackage{graphicx}

\usepackage{inputenc}
\usepackage{csquotes}
\usepackage[american]{babel}
\usepackage[T1]{fontenc}
\usepackage{enumerate}
\usepackage{mdwlist}
\usepackage[activate=normal]{pdfcprot}
\usepackage{bbding}
\usepackage{color}
\usepackage{amssymb}
\usepackage{amsmath}
\usepackage{amsfonts}
\usepackage{mathrsfs}

\usepackage{bm}
\usepackage{dcolumn}
\usepackage{color}
\usepackage[colorlinks=true,citecolor=blue]{hyperref}
\hypersetup{colorlinks=true,citecolor=blue,linkcolor=red
,urlcolor=blue}

\begin{document}

\title{Effect of boron and phosphorus codoping on the electronic and optical properties of graphitic carbon nitride monolayers: First-principle simulations}

\author{Mahdieh Yousefi}
\affiliation{Institute for Nanoscience and Nanotechnology, Sharif University of Technology, P.O. Box 14588-8969, Tehran, Iran}
\author{Monireh Faraji}
\affiliation{Chemistry Faculty, North Tehran Branch, Islamic Azad University, P.O.Box 16511-53311, Tehran, Iran}
\author{Reza Asgari}
\affiliation{School of Physics, Institute for Research in Fundamental Sciences (IPM), Tehran 19395-5531, Iran}
\affiliation{School of Nano Science, Institute for Research in Fundamental Sciences (IPM), Tehran 19395-5531, Iran}
\author{Alireza Z. Moshfegh}
\email{moshfegh@sharif.edu}
\affiliation{Department of Physics, Sharif University of Technology, P.O. Box 11555-9161, Tehran, Iran}
\affiliation{Institute for Nanoscience and Nanotechnology, Sharif University of Technology, P.O. Box 14588-8969, Tehran, Iran}

\begin{abstract}
We study the effect of boron (B) and Phosphorous (P) co-doping on electronic and optical properties of graphitic carbon nitride (g-C$_3$N$_4$ or GCN) monolayer using density functional simulations. The energy band structure indicates that the incorporation of B and P into a hexagonal lattice of GCN reduces the energy band gap from $3.1$ for pristine GCN to $1.9$ eV, thus extending light absorption toward visible region. Moreover, on the basis of calculating absorption spectra and dielectric function, the co-doped system exhibits an improved absorption intensity in the visible region and more electronic transitions, which named $\pi^*$ electronic transitions that occurred and were prohibited in the pristine GCN. These transitions can be attributed to charge redistribution upon doping, caused by distorted configurable B/P co-doped GCN confirmed by both electron density and Mulliken charge population. Therefore, B/P co-doped GCN is expected to be an auspicious candidate to be used as a promising photoelectrode in Photoelectrochemical water splitting reactions leading to efficient solar H$_2$ production.
\end{abstract}

\maketitle

\section{Introduction}
A global concern has been arisen owing to rapid industrial development and population growth, resulting in energy scarcity and earth pollution. In this regard, developing green and sustainable methods for producing clean energy and solving environmental pollution problems have absorbed enormous attention~\cite{abe2010recent, samadi2016recent}. Among various auspicious strategies, semiconductor photocatalysis has been widely studied in recent years owing to its capabilities to obtain hydrogen as an energy carrier, to remove organic pollutants, and to reduce CO$_2$ by converting solar energy into chemical energy. Therefore, the performance of photocatalytic materials is greatly dependent on the efficiency of visible light absorption because approximately 50 percent of sunlight consists of the visible part ~\cite{abe2010recent, li2013photoelectrochemical, ebrahimi2018facile}.

Recently, a metal-free semiconductor photocatalyst based on graphitic carbon nitride, g-C$_3$N$_4$, has received much attention \cite{PhysRevB.50.10362, PhysRevB.64.235416, PhysRevB.73.125427, wei2013strong} from a photocatalytic perspective because of its high thermal stability, chemical stability, and visible light absorption ~\cite{ding2016does}. However, pure g-C$_3$N$_4$ displays a poor photocatalytic efficiency owing to the low surface area, high recombination rate of photogenerated electron-hole pairs, and poor optical absorption above 420 nm \cite{zhang2015origin, naseri2017graphitic}. To avoid these drawbacks and enhance the photocatalytic performance, many attempts have been pursued to achieve a reasonable efficiency, such as exfoliating layered GCN into nanosheets \cite{xu2013chemical}, incorporation of foreign elements and impurities including Fe~\cite{wang2009metal}, Na~\cite{xiong2016bridging}, K~\cite{xiong2016bridging} ,Li~\cite{zhu2014lithium}, P~\cite{guo2016phosphorus}, O~\cite{bu2014effect}, N~\cite{zhou2016n}, coupling with metals such as Ag~\cite{bai2014enhancement}  and Au~\cite{cheng2013nanoparticle}, inorganic semiconductors like TiO$_2$~\cite{chen2016heterojunctions}, and layered semiconductors such as MoS$_2$~\cite{ge2013synthesis}.

To improve the charge transfer kinetics in GCN, many efforts have put into designing and constructing GCN-based heterojunction using different semiconductors, such as Fe$_2$O$_3$ and TiO$_2$, with a proper valence band and conduction band  potentials ~\cite{ong2016graphitic}. Considering the GCN-nanosheets-based heterojunction, composed of two components and three heterostructures~ \cite{xu2015sulfur, li2016novel, yan2016construction} with different types of interfaces have been propounded so as to enhance the photocatalytic performance of GCN. Furthermore, the incorporation of GCN with metals (particularly noble metals such as Au and Ag) is an effective way to exploit the charge kinetics of GCN \cite{ong2016graphitic}. In this regard, upon light absorption, collective oscillations of free electrons, known as localized surface plasmon resonance effect, occurs. This surface plasmon results in extending light absorption substantially into the visible and thus increasing the number of photogenerated electron-hole pairs in the adjacent semiconductor. Additionally, the incorporation of foreign elements and impurities into the GCN framework is an intriguing way to promote the electrical, optical, and surface properties of GCN \cite{ong2016graphitic}.

Co-doping is a promising technique that can be used for effectively tuning the dopant populations, electronic and optical properties. It can enhance the solubility of dopants and improve the stability of desired defects. Recently, Zhang et al.~\cite{zhang2010phosphorus} have investigated the effect of P doping on electrical characteristic of g-C$_3$N$_4$. Their results indicated that electrical conductivity increases remarkably upon phosphorous doping, leading to a higher charge carrier density. Furthermore, Sagara et al.~\cite{sagara2016photoelectrochemical} have found that B doped GCN electrode shows a far better CO$_2$ reduction activity than that of pure g-C$_3$N$_4$ electrode. To do so, boron and phosphorous co-doped g-C$_3$N$_4$ has been experimentally reported by Razig et al.~\cite{raziq2017synthesis} very recently. On the basis of their report, the optimized nanocomposites exhibit improved visible-light activities for CO$_2$ conversion as well as phenol and acetaldehyde degradation. Additionally, Srinivasu et al.~\cite{srinivasu2014porous} theoretically illustrated that incorporation of P or B elements into g-CN, another form of graphitic carbon nitride with CN stoichiometry, enhanced the charge carrier mobility. It is worth mentioning that in the last five years, various studies have been conducted on photocatalytic properties of GCN as a two-dimensional (2D) layered system synthesized via chemical, liquid, ultrasound, and thermal exfoliation. Synthesizing GCN nanosheets could be attributed to the fact that bulk GCN possesses a high degree of grain boundary defects due to preparation at high temperatures resulting in high electron-hole recombination rate. Hence, by exfoliating layered GCN into nanosheets, improved electronic properties and high specific surface area could be achieved.

In this work, GCN monolayer is chosen to be doped with phosphorous and boron impurities. Using first principles based calculations, we study the geometries, electronic, and optical properties of B doped and P doped as well as B/P co-doped g-C$_3$N$_4$ monolayer and their results are compared. We show that the incorporation of both B and P into a hexagonal lattice of GCN reduces the energy band gap from $3.1$ for pristine GCN to $1.9$ eV. Moreover, the co-doped system exhibits an improved absorption intensity in the visible region and more electronic transitions which are prohibited in the pristine GCN. Therefore, B/P co-doped GCN is expected to be an promising material to be used in many chemical and optical applications.

This paper is organized as follows. In Sec.~\ref{sec:model}, we introduce our system and model and explain the method which is used to calculate the electronic and optical properties of the system. In Sec.~\ref{sec:results}, we present and describe the numerical results for the  of a GCN reside on a co-doping effects. Finally, we conclude and summarize our main results in Sec.~\ref{sec:concl}.

\begin{figure*}[t]
\centering
  \includegraphics[width=15cm,height=18cm]{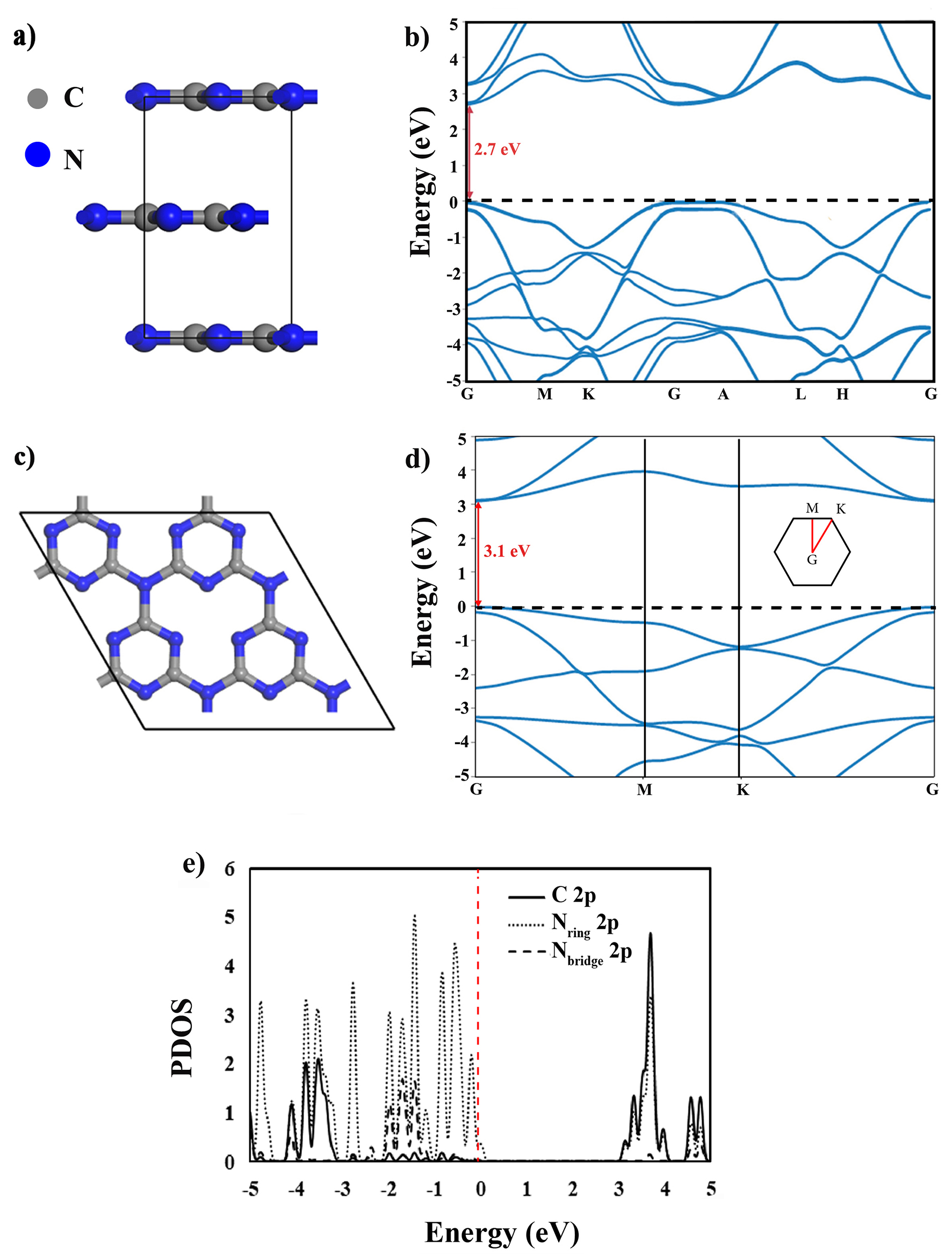}
  \caption{ (Color online) (a) Optimized structure of a GCN unit cell with hexagonal layered structure and (b) band structure of bulk GCN which indicates a semiconducting behavior with a direct band gap energy of $2.70$ eV, (c) optimized ($2\times$2) supercell, (d) band structure which is  $3.11$ eV, and (e) PDOS calculations of GCN monolayer. Notice that the upper valence band composes of the p states of N$_{\rm ring}$, while the lowest conduction band prevailingly originates from p states of C and N$_{\rm ring}$ atoms. Moreover, some electrons, extracted from the C atoms, delocalized over the N atoms. The $1\times 1$ supercell is used here.}
  \label{fig1}
\end{figure*}

\section{Methods and Computational details}\label{sec:model}
In this study, the density functional theory (DFT) simulations are carried out by using the CASTEP code~\cite{segall2002first}. All crystalline cells, including bulk GCN, monolayer of GCN, B and P-doped, and B/P co-doped GCN are optimized within the generalized gradient approximation (GGA) and the exchange-correlation functional of the Perdew-Burke-Ernzerhof (PBE). To consider the ion-electron interactions, the ultrasoft pseudo-potential is employed. The kinetic cutoff energy of $500$ eV and sampling of the reciprocal space Brillouin zone is done by a Monkhorst-pack grid of $6\times6$ $k$-point to perform geometry optimization and electronic structure calculations. A vacuum slab of $20$ $\mbox{\AA}$ along the $z$ direction (normal to the GCN monolayer) is used to all the pure and doped monolayer systems to avoid the interaction between neighboring cells. All atomic positions and lattice parameters are allowed to relax until the convergence threshold for energy, the maximum Hellmann-Feyman forces on each atom, and the maximum displacement are less than $1.0\times10^{-5}$ Hartree atomic units , 0.002 Hartree atomic/\AA, and 0.005 \AA, respectively. It should be noted that to obtain more accurate calculations, a HSE06 hybrid functional is also employed. Moreover, we will consider diluted doping concentrations, therefore, the defect-defect interaction should be very small and we ignore this effect in our calculations.

\section{Results and discussion}\label{sec:results}
\subsection{GCN bulk monolayer}
In this section, we present our main numerical results based on first-principles simulations. Our aim is to explore the impact of co-doping on electronic and optical properties of GCN. All the first-principles calculations are performed at zero temperature.

Figure \ref{fig1}(a) presents the unit cell of triazine bulk GCN with hexagonal layered structure, belonging to space group P = 6m2 (No. 187). The calculated optimized lattice constants, $a = b = 4.81$ and $c = 6.27$ \AA are in good agreement with experimental measurements and theoretical studies \cite{Liu2016, sun2008solvent}. The calculated band structure of bulk GCN is plotted in Fig. \ref{fig1}(b). Clearly, bulk GCN indicates a semiconducting behavior with a direct band gap energy of $2.70$ eV, which is consistent with obtaining experimental band gaps \cite{yang2013exfoliated}. Considering the GCN monolayer, shown in Fig. \ref{fig1}(c), two kinds of nitrogen atoms, namely N$_{\rm ring}$ and N$_{\rm bridge}$, are observed because of different chemical environments~\cite{Liu2016}. In fact, although N$_{\rm bridge}$ atoms are fully saturated by three surrounding C atoms, N$_{\rm ring}$ atoms only connect two C atoms, leaving a non-bonding character behind~\cite{xu2015insights}. In this regard, two kinds of bond lengths are calculated around 1.47 and 1.33 \AA for C-$N_{\rm bridge}$ and C-$N_{\rm ring}$, respectively. The calculated band gap of GCN monolayer is $3.11$ eV (Fig.~\ref{fig1}(d)), which is in reasonable agreement with the HSE06 band gap of $3.18$ eV~\cite{cui2015structural}. Moreover, the different chemical bonding environments of nitrogen atoms are also confirmed by the calculated partial density of states (PDOS), as shown in Fig. \ref{fig1}e. On the basis of PDOS calculations, the upper valence band composes of the p states of N$_{\rm ring}$, while the lowest conduction band prevailingly originates from p states of C and N$_{\rm ring}$ atoms~\cite{xu2015insights}. Moreover, since nitrogen is more electronegative than carbon, some electrons, extracted from the C atoms, delocalized over the N atoms~\cite{ma2012strategy}. This behavior is confirmed by Mulliken population analysis in which the electron distributions at N$_{\rm ring}$, N$_{\rm bridge}$, and C are -0.410, -0.330, and +0.520 electron, respectively. It would be worth mentioning that the Kohn-Sham eigenvalues do not correspond, in general, to physical excitation energies of the system and therefore the PDOS, stemming from DFT, provides
a qualitative picture of the accurate PDOS of the system.

Table \ref{tab1} presents some reported band gap values obtained experimentally as well as theoretically using DFT calculations based on implication of different exchange correlation functionals. It should be noted that the calculated band gap of the pure GCN is always underestimated by generalized gradient approximations~\cite{wei2013strong}. The increment in band gap value for the GCN monolayer can be assigned to the quantum confinement effect.

 \begin {table}[b]
  \caption{The calculated and experimental band gap energy of pure and doped GCN}
  \begin{center}
    \begin{tabular}{|c|c|p {1.6 cm}|p {1.3 cm}|p {1 cm}|}
  \hline
 Structure &	Method	& Exchange correlation function & Band gap (eV)	&Ref.\\
 \hline
 GCN monolayer &	Experimental & - &	2.92  &\cite{xu2013chemical} \\
 \hline
 Layered GCN &	Experimental &	- & 2.79  & \cite{lin2015efficient}\\
   \hline
   Layered GCN &	Experimental & - &	2.82  &\cite{ma2016water} \\
   \hline
   GCN monolayer & Theoretical & HSE06 &	3.03 & \cite{zhang2015origin}\\
   \hline
  GCN monolayer	& Theoretical & HSE06 &	3.18 & \cite{cui2015structural}\\
  \hline
  Layered GCN &	Theoretical & LDA	& 1.43  &\cite{wei2013strong}\\
  \hline
  Layered GCN &	Theoretical & GGA	& 1.60 & \cite{gao2016atomically}\\
  \hline
    P doped GCN &	Theoretical & HSE06 & 2.01 &	\cite{ma2012strategy} \\
        \hline
        P doped GCN  & Theoretical & HSE06 & 2.55 &	\cite{srinivasu2014porous} \\
        \hline
      B doped GCN  & Theoretical & HSE06 & Half metallic &	\cite{meng2015half} \\
       \hline
         GCN monolayer &	Theoretical & HSE06 &	3.10 &	Present work \\
              \hline
       B doped GCN & Theoretical & HSE06 &	Metallic &	Present work \\
              \hline
       P doped GCN & Theoretical &HSE06 & Half-filled metalic &	Present work \\
                            \hline

  \end{tabular}
  \label{tab1}

 \end{center}
  \end{table}

 \begin{figure}[t]
   \centering
     \includegraphics[width=3.3 in]{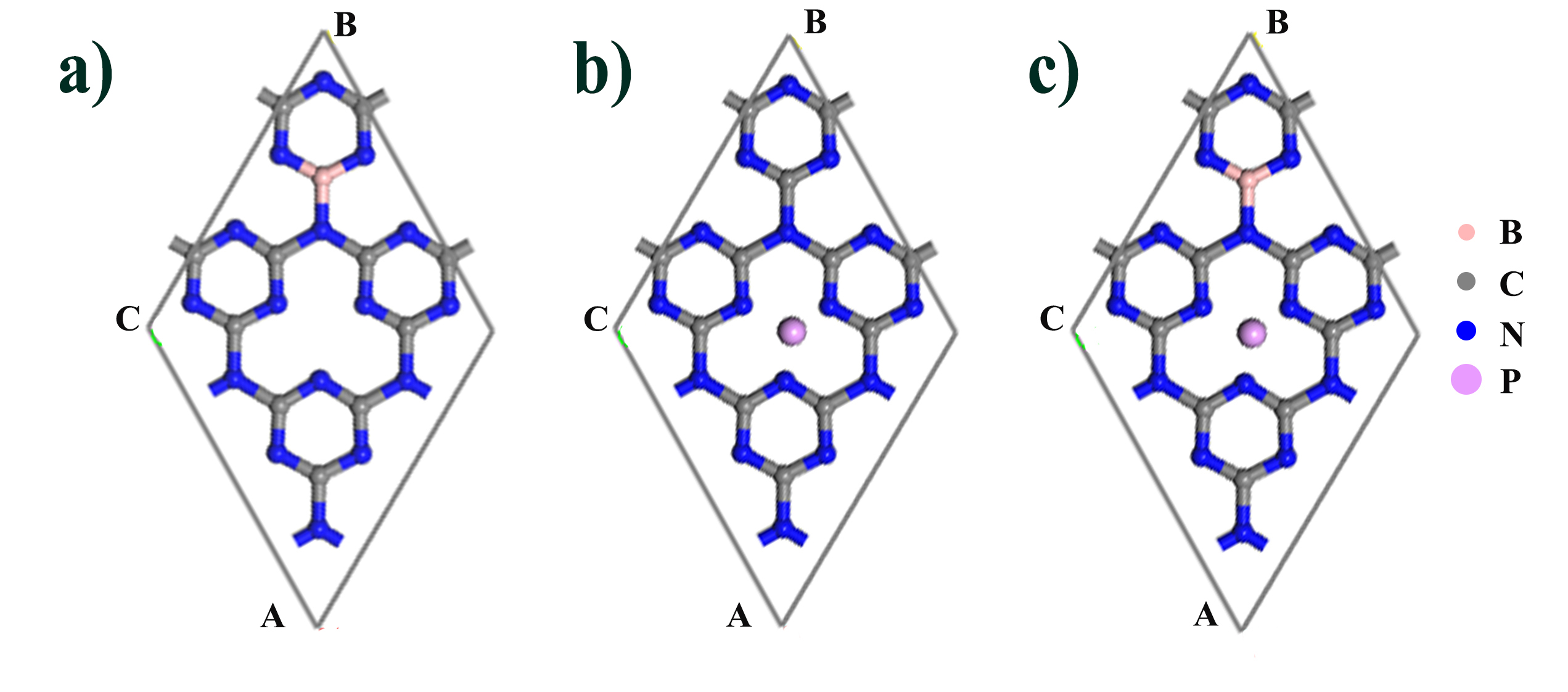}
     \caption{ (Color online) The front views of the (a) B doped, (b) P doped, and (c) B/P co-doped GCN monolayer. }
   \label{fig2}
   \end{figure}

  \subsection{Doped GCN monolayer}
 In order to dope GCN with both boron and phosphorous atoms, four possible sites, including C, N$_{\rm ring}$, N$_{\rm bridge}$, and interstitial, can be considered. In the case of boron, substitution of B for C atoms is energetically most favorable~\cite{ding2016does}. For phosphorous doping, it is found that P atom cannot substitute the carbon or both types of nitrogen atoms. However, interstitially P doped GCN is reported to be the most stable configuration thermodynamically~\cite{wen2017review}. Finally for the B/P co-doped system, B atom substitutes the C atom and P atom chemically adsorbs on the GCN monolayer as shown in Fig. \ref{fig2}(a), \ref{fig2}(b), and \ref{fig2}(c).

 \begin{figure}[t]
    \centering
      \includegraphics[width = 3.1 in]{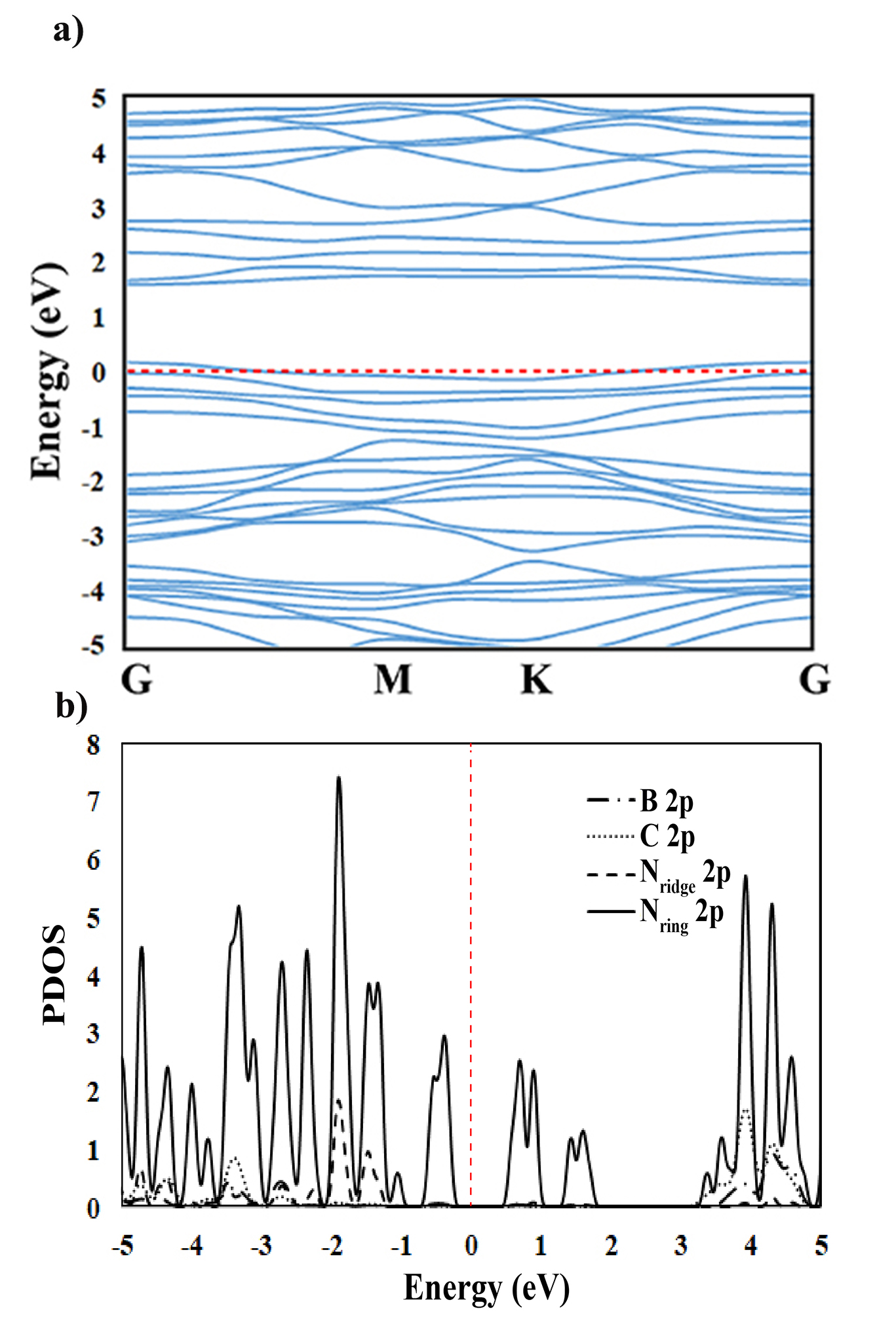}
       \caption{(Color online) (a) Band structure and (b) PDOS calculations of B-doped GCN monolayer. A metallic behavior is induced in the GCN monolayer when a
       C atom is replaced by the B atom. The metallicity is principally dominated by p orbitals of N$_{\rm ring}$ atoms, which are connected with the B atom.
       N$_{\rm bridge}$ and boron atoms have a little contribution as the PDOS illustrates in (b).
       For the $2\times2$ supercell, there are 28 atoms. For the boron doped GCN, the concentration is $3.57$\%.
       In the case of interstitial doped GCN the concentration is $3.44$\%.
       Finally, for the co-doped system, the concentration is $6.89$\%. }
    \label{fig3}
    \end{figure}

     \begin{figure}[t]
          \centering
            \includegraphics[width = 3.1 in]{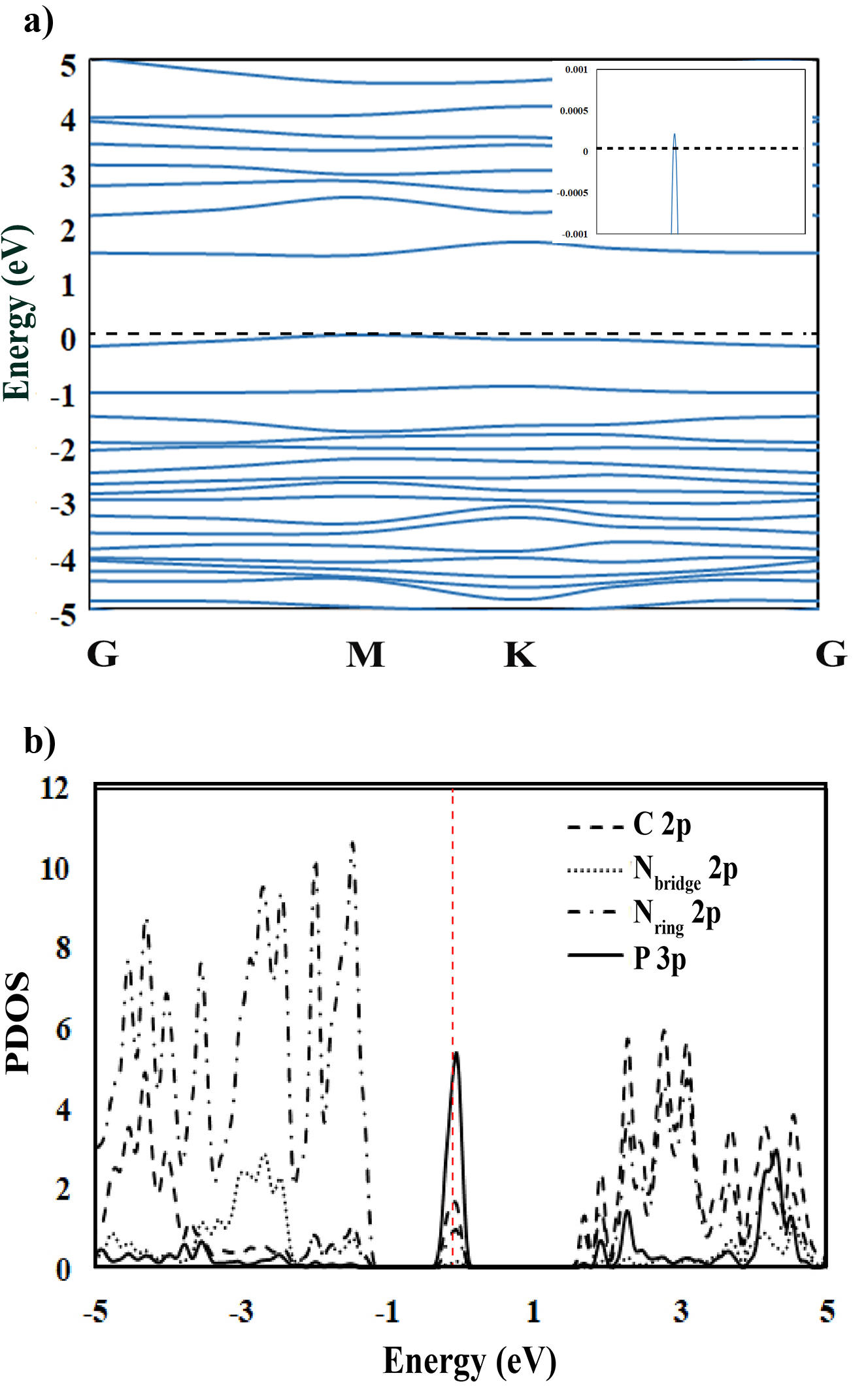}
                \caption{(Color online) (a) Band structure and (b) PDOS calculations of the P-doped GCN monolayer. A half-filled metalic behavior is found. Inset: Small intersection between the band structure and the Fermi energy is shown. Notice that the PDOS of P interstitially doped GCN is remarkably different from that of the un-doped system. Moreover, doping GCN with the P atom leads to a reduction in the contribution of p states of N$_{\rm ring}$ atom to the valence band edge, whereas there is a little increment in the contribution of p states on N$_{\rm bridge}$ atom to the conduction band edge. Besides, the P atom prefers to bind two adjacent N$_{\rm ring}$ atoms. The $2\times 2$ supercell is used here.}
          \label{fig4}
          \end{figure}

To study the stability of the mono-doped system, the formation energy ($E_f$) can be calculated using the following equations:
\begin{eqnarray}
     E_{f}&=& E_{T}(sub)-E_{T}(pure)-\mu_{A}+\mu_{B}\nonumber\\
        E_{f}&=& E_{T}(int)-E_{T}(pure)-\mu_{A}
\end{eqnarray}
where $E_{T}(pure)$, $E_{T}(sub)$, and $E_{T}(int)$ are the total energies of the pristine GCN, doped GCN for substitutional and interestitial dopants, respectively. $\mu_{A}$ and $\mu_{B}$ are the chemical potentials of the phosphorous (boron) and carbon/nitrogen atoms, respectively. For GCN, the relation $3\mu_{C}+4\mu_{N}=\mu(GCN)$ should be satisfied. To determine the chemical potentials, solid graphite\cite{ding2016does, ma2012strategy, Liu2016}, boron nitride (BN) \cite{ding2016does}, and $P_{4}$ are used \cite{ma2012strategy, Liu2016}: $\mu_{C}=\frac{\mu(graphite)}{4}$, $\mu_{N}=\frac{(\mu(graphite)-3\mu)}{4}$, $\mu_{B}=\mu(BN)-\mu_{N}$, and $\mu_{P}=\frac{\mu(P_4)}{4}$ \cite{franck1990jd, wiberg1972chemische, chase1974janaf}. In the case of triazine GCN monolayer, there are two inequivalent N sites, N$_{\rm ring}$ and N$_{\rm bridge}$, and all carbon atoms are chemically equivalent. The dopant formation energies of B doped and P doped GCN monolayer are reported in Table 2. As seen from \ref{tab2}, the B substituted carbon and P interstitially doped systems indicate the lowest formation energies. Therefore, these systems are energetically more favorable to form.
\begin {table}[ht]
  \caption{Dopant Formation Energies (eV) of GCN monolayer}
  \centering

    \begin{tabular}{|c|c|c|c|c|c|c|c|}
  \hline
 B$_{\rm Nring}$ & B$_{\rm Nbridge}$	& B$_{\rm C}$ & B$_{\rm i}$ & P$_{\rm Nring}$& P$_{\rm Nbridge}$ & P$_{\rm C}$ & P$_{\rm i}$\\
 \hline
 3.95 &	3.25 &	1.35  & 1.95 & 3.52 & 0.83 & 1.42 & 0.75 \\
 \hline

  \end{tabular}

  \label{tab2}

    \end{table}

\begin{figure}[t]
        \centering
          \includegraphics[width = 3.1 in]{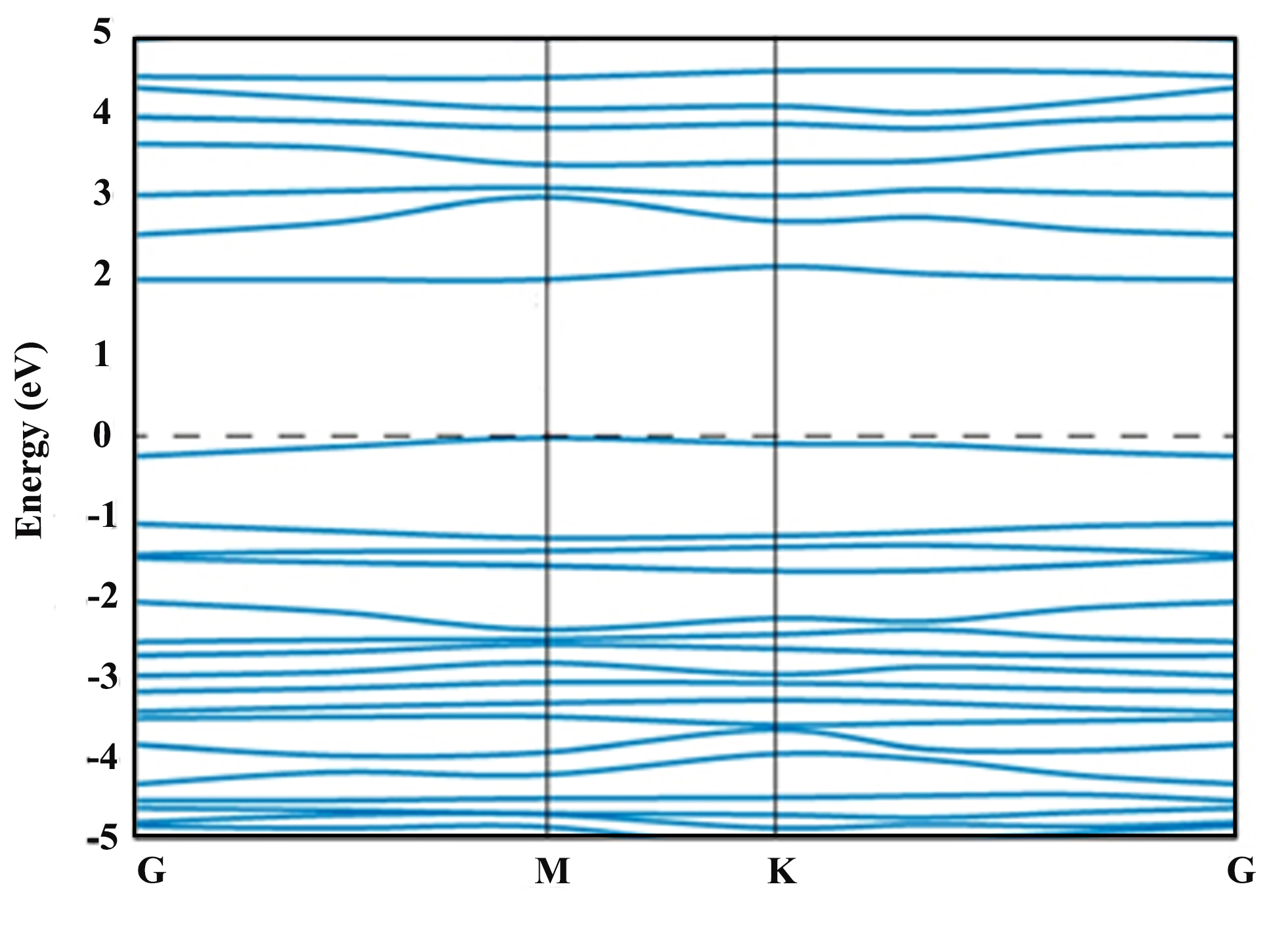}
          \caption{(Color online) The band structure of the B/P co-doped GCN monolayer. The band gap is almost 1.95 eV. The $2\times 2$ supercell is used here. }
        \label{fig5}
        \end{figure}
\begin{figure}[t]
               \centering
                 \includegraphics[width = 3.3 in]{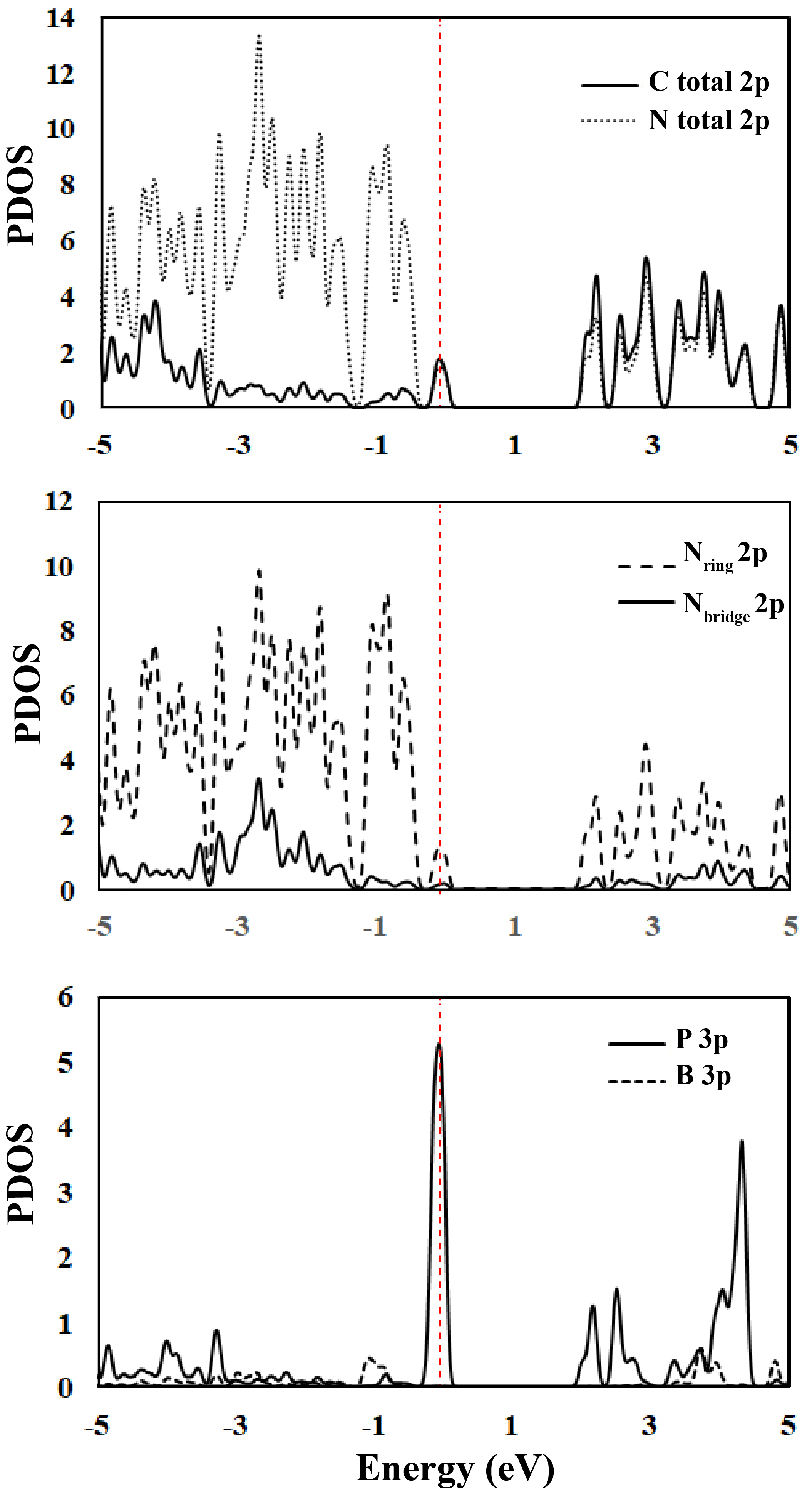}
                  \caption{(Color online) PDOS calculations of the B/P co-doped GCN monolayer. The contribution of p states of N$_{\rm ring}$ atoms reduces for the B/P co-doped GCN. Moreover, the p states of phosphorous and carbon have contributions to the both conduction and valence bands, and electrons can be excited from P and C atoms. Two P-N$_{\rm ring}$ bond lengths are almost similar to those of the P doped GCN. The bond length of B atom with two adjacent N$_{\rm ring}$ atoms are slightly different for the co-doped system. This difference can be assigned to the severe deformation of the planar shape of GCN monolayer when P is added to the system. }
               \label{fig6}
               \end{figure}
               \begin{figure}[t]
         \centering
           \includegraphics[width = 3.1 in]{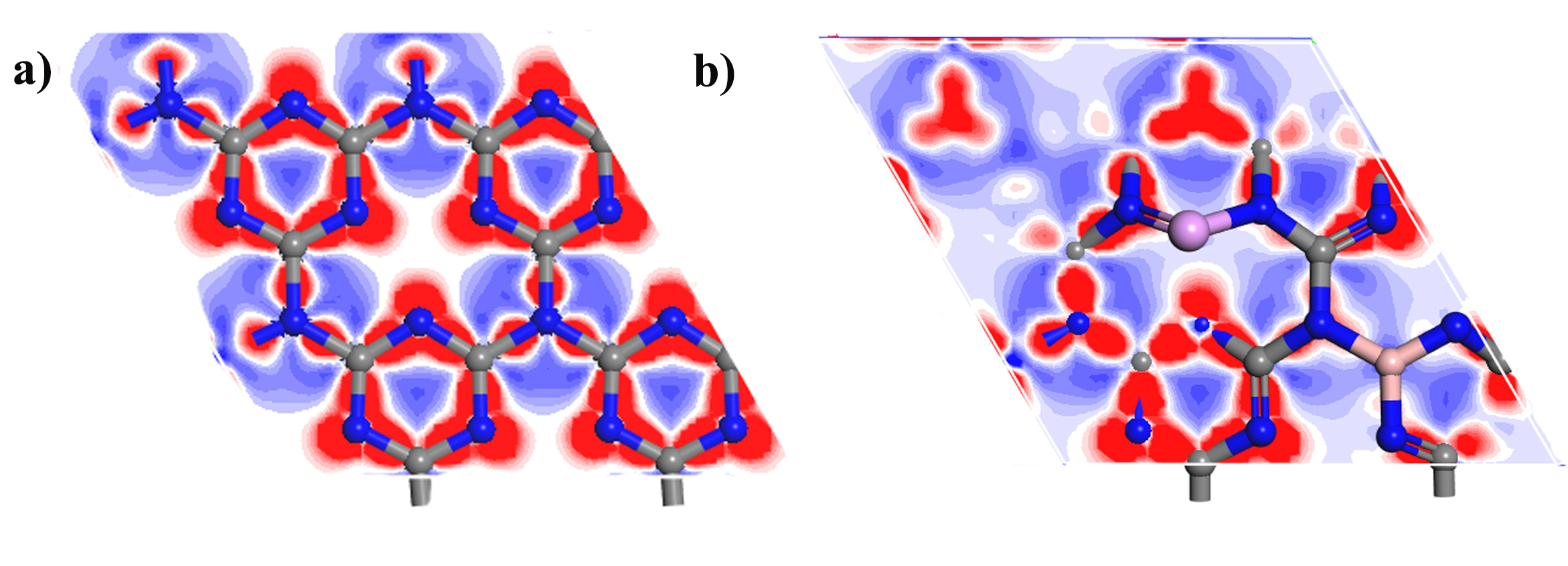}
           \caption{(Color online) Difference charge density contour maps projected on the parallel plane for (a) pristine and (b) B/P codoped GCN monolayer. }
         \label{fig7}
         \end{figure}
 \begin{figure*}[t]
 \centering
  \includegraphics[width=15cm,height=6cm]{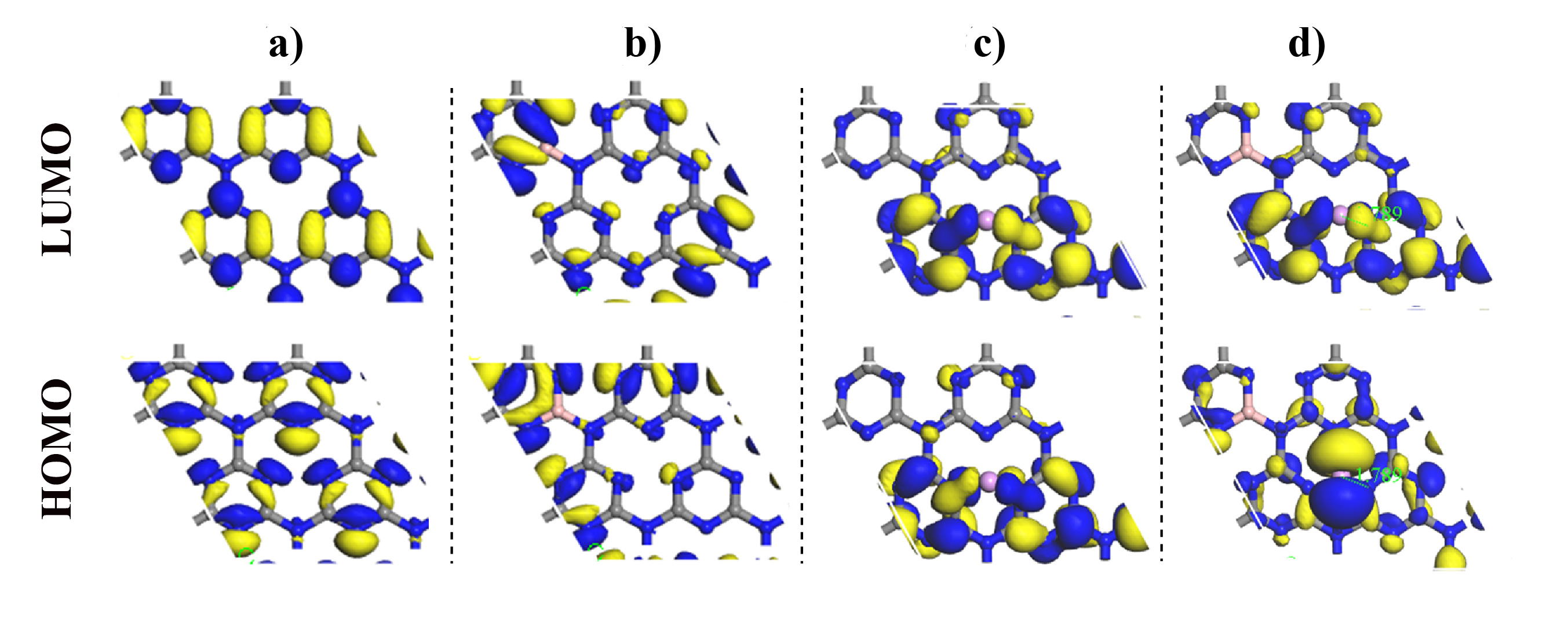}
  \caption{(Color online) LUMO and HOMO of (a) pure , (b) B doped, (c) P doped, and (d) B/P codoped GCN monolayer. The Fermi level is set to the zero of energy. Gray and blue spheres represent the C and N atoms, respectively. }
  \label{fig8}
\end{figure*}
We show the electronic band structure of B doped GCN monolayer in Fig. \ref{fig3}(a). Our numerical calculation show that a metallic behavior is induced in the GCN monolayer when a C atom is replaced by the B atom. The metallicity is principally dominated by p orbitals of N$_{\rm ring}$ atoms, which are connected with the B atom, nevertheless N$_{\rm bridge}$ and boron atoms have a little contribution as the PDOS illustrates in Fig. \ref{fig3}(b). The lattice parameter, $a$, increased to $4.99$ {\AA} upon doping with the B atom. Furthermore, the lengths of B-N$_{\rm ring}$ and B-N$_{\rm bridge}$ bonds are 1.45 and 1.51 \AA, respectively, which are higher than C-N$_{\rm ring}$ and C-N$_{\rm bridge}$ bonds in the pure GCN monolayer. The formation of the weaker covalent bonds between the B atom and the adjacent N$_{\rm ring}$ and N$_{\rm bridge}$ atoms may stem from the smaller absolute electronegativity of the B atom (4.29) than those of the C atom (6.27) and the N atom (7.30) on the Pauling scale~\cite{ma2012strategy}. Mulliken population analysis suggests that nitrogen atoms gained -0.150 and -0.260 electron, however, the C atom lost +0.060 electron. Hence, the electron distribution at Nring, Nbridge, and C becomes -0.560, -0.590, and +0.580 electron, respectively. This charge redistribution at the N and C atoms produces an electric field near the surface of the GCN monolayer. Despite the fact that the charge redistribution may favor the charge carrier separation, the observed metallic behavior is not appropriate to be utilized in photocatalytic applications. It should be noted that a half-metallic behavior was observed by Gao et al.~\cite{gao2016atomically} for B doped heptazine GCN prepared by heating the mixture of melamine and boron oxide.

For the P-doped GCN monolayer, a metallic behavior is observed. The band structure is shown in Fig. \ref{fig4}(a). Basically, the Fermi level crosses the top of the highest valence band in a small window in k-space. Since an appropriate band gap energy higher than 1.9 eV must be used in any photocatalytic applications, this system cannot be utilized as a visible photocatalyst. To get further insight into the electronic structures of the P doped GCN monolayer, its PDOS is illustrated in Fig. \ref{fig4}(b). As shown, the PDOS of P interstitially doped GCN is remarkably different from that of the un-doped system. Doping GCN with the P atom leads to a reduction in the contribution of p states of N$_{\rm ring}$ atom to the valence band edge, whereas there is a little increment in the contribution of p states on N$_{\rm bridge}$ atom to the conduction band edge. The p states of P and C atoms contribute both the valence and conduction band edges of the P doped GCN, and, as a result, electrons in the valence band edge can be excited from the P and C atoms. Although B doped system conserves its planar structure after optimization, the original planar shape is broken in the P doped system, resulting in the deformation of the overall $\pi$-conjugated electronic states in the triazine unit. Moreover, according to the obtained results, the P atom prefers to bind two adjacent N$_{\rm ring}$ atoms. Two P-N$_{\rm ring}$ bond lengths are almost similar and are calculated around 1.78 \AA, which are weaker than other C-N$_{\rm ring}$ bonds. The Mulliken charge population analysis also shows that the P atom loses electrons and is positive in charge (+0.430 electron). It should be noted that for P doped GCN configuration $sp^2$ hybrid orbitals of phosphorus bonded with two $sp^2$ hybrid orbitals of the adjacent nitrogen, while a lone-pair electrons localize around the P atom and an electron delocalizes around NPN chain.

\begin{figure*}[t]
 \centering
  \includegraphics[width=15cm,height=12.5cm]{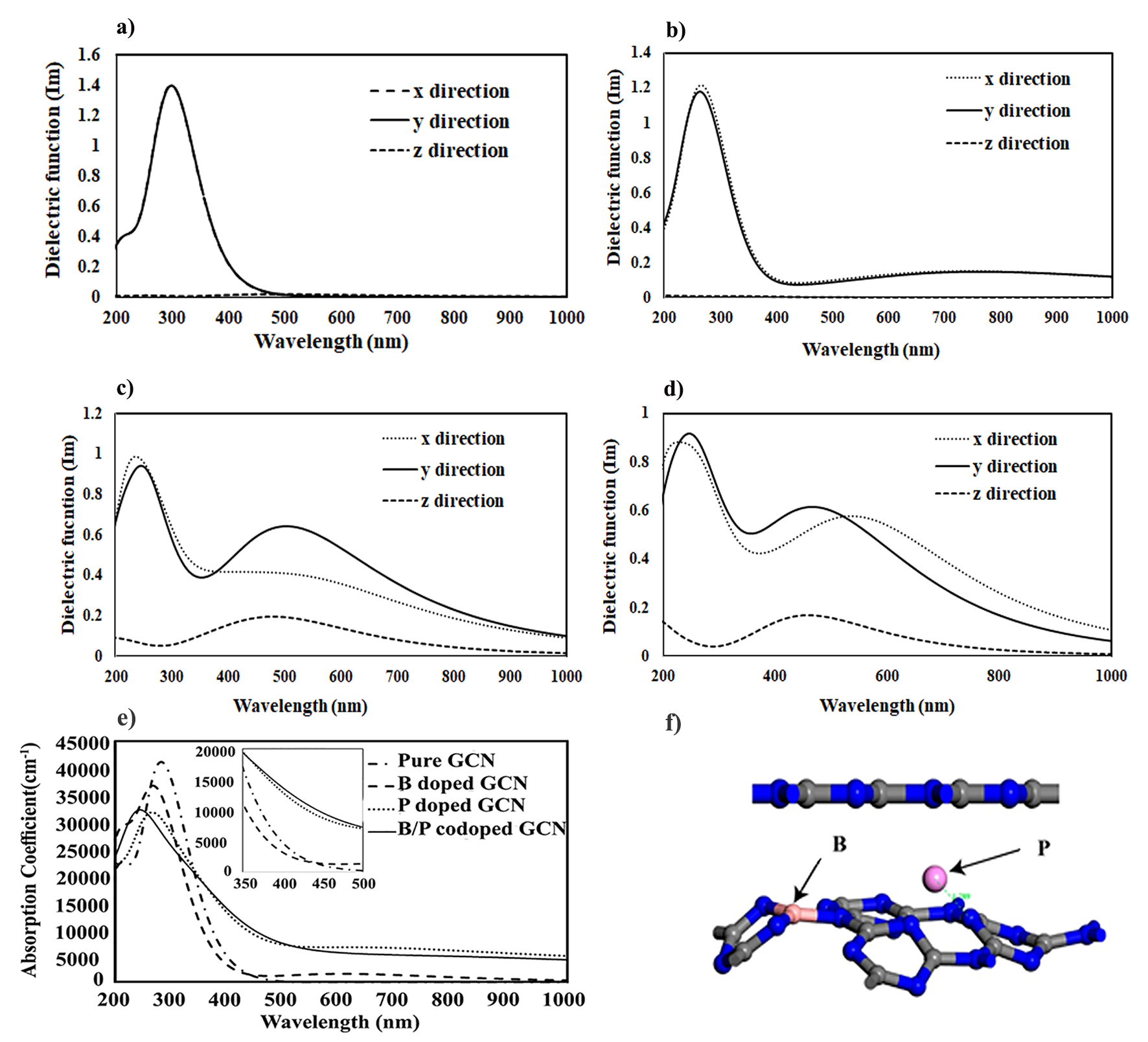}
      \caption{(Color online) Imaginary part of the dielectric function (a) Pure, (b) B-doped. (c) P-doped and (d)B/P codoped GCN monolayer. (e)  absorption coefficient of pure and doped GCN monolayer, and (f) the side views of pristine and B/P codoped GCN. Gray and blue represent C and N atoms. The imaginary part of the dielectric function depends of the light polarization. The inset in (e) is zoom for a finite wavelength. The pure GCN monolayer shows a strong absorption peak around 270-320 nm, attributed to the $\pi$-$\pi^*$ electronic transition. Notice that the B/P co-doped GCN monolayer improves the utilization of visible portion of solar irradiation observed by experiment~\cite{raziq2017synthesis}. In addition, the dielectric function depends on the
material density, the interlayer distance and the type of doped system.  }
           \label{fig9}
           \end{figure*}

In the case of B/P co-doped GCN monolayer, the band gap is determined of around 1.95 eV which is higher than that of the P doped GCN and our numerical results are shown in Fig. \ref{fig5}. This increment in band gap energy from 1.43 to 1.95 eV makes the co-doped system an appropriate candidate to be used in photocatalytic applications for water splitting. It is worth mentioning that the bandwidths are smaller for the P doped and B/P co-doped GCN monolayer than that for the pure or a B doped system. Therefore, the mobility of charge carriers decreases. In spite of the reduction in the mobility of charge carriers, which may be compensated by migration of excitons through a GCN single layer, doping of GCN with B and P can be an effective strategy so as to enhance the optical properties of this structure.

The PDOS plots of the co-doped system is indicated in Fig. \ref{fig6}. Similar to the P doped GCN, the contribution of p states of N$_{\rm ring}$ atoms reduces for the B/P co-doped GCN. Moreover, the p states of phosphorous and carbon have contributions to the both conduction and valence bands, and, as a result, electrons can be excited from P and C atoms. As seen, unlike the P atom, the B atom has no contribution to the conduction and valence band edges. Two P-N$_{\rm ring}$ bond lengths are almost similar to those of the P doped GCN, however, the band lengths of B-N$_{\rm ring}$ and B-N$_{\rm bridge}$ are slightly different from those of B doped GCN. Furthermore, the lattice parameters of $a$ and $b$ vary around 2.9 and 2.7 percent, respectively, for the co-doped system. It should be noted that unlike the B doped system, in which the lengths of two B-N$_{\rm ring}$ bonds are similar, the bond length of B atom with two adjacent N$_{\rm ring}$ atoms are slightly different for the co-doped system. This difference can be assigned to the severe deformation of the planar shape of GCN monolayer when P is added to the system. Furthermore, the Mulliken charge of P and B atoms in co-doped configuration are +0.51 and +1.06 electron which differs from those of B and P doped systems.

In the case of charge distribution over the pure GCN monolayer, a homogeneous electron density difference, projected on the parallel plane contains the system layer, owing to the formation of a sinusoidal-like shape is shown in Fig. \ref{fig7}(b). This distortion leads to charge redistribution and, as a result, an internal electric field may form leading to the retardation of charge recombination.
\begin{figure}[h]
                     \centering
                        \includegraphics[width = 3.3 in]{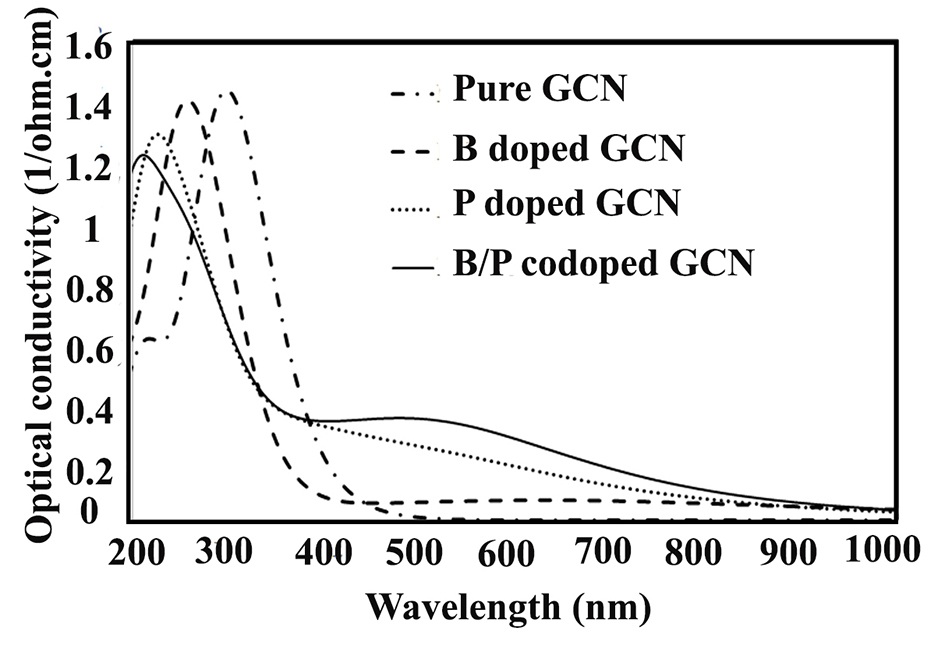}
                        \caption{(Color online)  optical conductivity for pure and doped GCN monolayer. Note that doped system exhibits more effective ultraviolet absorption and enhanced visible-light response. This behavior is also confirmed by our optical conductivity calculations. A rather sharp increase in the optical conductivity takes place in the ultra-violate region and the correspondent peak moves toward lower wavelength with B, P and B/P doping.   }
                     \label{fig10}
                      \end{figure}

To investigate the variation of frontier orbitals induced by the addition of B, P, and both B/P into GCN monolayer, the highest occupied molecular orbital (HOMO) and lowest unoccupied molecular orbital (LUMO) are illustrated in Fig. \ref{fig8}. It is well-known that pure GCN is $\pi$-delocalized with $sp^2$ hybridization \cite{li2017mechanistic}. According to Fig. \ref{fig8}(a), the LUMO is mainly dominated by 2p states of C and N$_{\rm ring}$ atoms, and, as a result, these atoms incline to provide reduction sites for H$^+$ to H$_2$ \cite{lu2017effects}, whereas the HOMO is mainly originated from 2p states of N$_{\rm ring}$ atoms preferring oxidation sites for H$_2$O to O$_2$~\cite{lu2017effects}. This result is in a good consistent with the PDOS calculations. It should be noted that since electrons and holes are separated on the neighboring C and N$_{\rm ring}$ atoms upon the light radiation, the recombination can be occurred easily. Therefore, GCN has a low photocatalytic efficiency under visible light irradiation. Upon the doping of GCN with B, P, and both B/P, a certain redistribution of HOMO and LUMO is observed. In the case of B doped GCN monolayer, the HOMO and LUMO mostly locate in the N$_{\rm ring}$ atom bonded with boron. Therefore, The LUMO and HOMO of B doped GCN are dominated by the 2p states of N$_{\rm ring}$ atom bonded with B (Fig. \ref{fig8}(b)). For P doped and B/P co-doped GCN monolayer, the formation of the sinusoidal-like structure results in weakening the $\pi$-delocalization, thus the intrinsic electron distributions vary as compared to pure GCN. According to the results shown in Fig. \ref{fig8}(c) and \ref{fig8}(d), for P doped and B/P co-doped systems, the LUMO and HOMO are great N$_{\rm ring}$, C, and P atoms. It is believed that the photogenerated charge carriers transfer freely by the way of C-N-P-N-C chain between two adjacent triazine units \cite{ma2012strategy}.  On the basis of partly or totally separated HOMO and LUMO of doped GCN monolayer, the lifetime of photogenerated electron-hole pairs as well the carrier mobility may greatly enhance.

 \subsection{Optical characteristic }
 In order to evaluate the performance of a photocatalyst in the visible-light region, it is very crucial to investigate its optical properties. In this context, the absorption spectra, the imaginary part of the dielectric function, and the optical conductivity of pure and doped systems are calculated using the Fermi golden rule within the dipole approximation by means of HSE06 functional. Measurement of the absorption of light is one of the most important techniques for optical
 measurements in solids. The absorption coefficient, $\alpha(\omega)=2\omega \kappa(\omega)/c$ where $\kappa(\omega)$ is the imaginary part of the complex index
 of refraction, is obtained from the following equation~\cite{srinivasu2014porous,fox2002optical}

 \begin{equation}
 \begin{split}
  \alpha(\omega)= \frac{\sqrt{2}\omega}{c}(\sqrt{\varepsilon_1(\omega)^2+\varepsilon_2(\omega)^2}-\varepsilon_1(\omega))^{\frac{1}{2}}
   \end{split}
\end{equation}
where $\varepsilon_1$ and $\varepsilon_2$ are the real and imaginary parts of the dielectric function, respectively. The real part of the dielectric function is obtained by a Kramers-Kronig transformation, and the imaginary part can be expressed by the following  \cite{saha2000structural},
 \begin{equation}
 \begin{split}
    \varepsilon_2(\omega)= \frac{Ve^2}{2\pi\hbar m^2\omega^2} \int d^3k \sum_{n,n'}|\langle kn|p|k'n\rangle|^2 f_n(k)\\
    (1-f_{n'}(k))\delta(E_n(k)-E_{n'}(k)-\hbar\omega)
     \end{split}
  \end{equation}

Here, $\hbar\omega$ is the energy of the incident photon, $m$ is the electron mass, p is the momentum operator, $|kn>$ is a crystal wave function and $f_n(k)$ is the Fermi distribution function with the energy $E_n(k)$ of band $n$. The dielectric function is calculated by invoking the Kohn-Sham wave functions and they change by changing the structures and inter-atomic interactions. Therefore, the dielectric function depends on the
material density, the interlayer distance and the type of doped system.

The absorption coefficient determines the fraction of energy lost by the electromagnetic wave when it penetrates through a unit thickness of the material~\cite{ma2012strategy}. Figures \ref{fig9}(a)-(b) indicate the imaginary part of the dielectric function of pristine and doped GCN monolayer in different light polarization. Compared with the GCN monolayer, doped system exhibits more effective ultra-violate absorption and enhanced visible light response. It can be seen from the curves of imaginary parts of the dielectric function, the dielectric spectrum under the polarization in $z$ direction is obviously different from those under $x$ and $y$ directions. This difference can be assigned to the symmetry of the dielectric spectra correspond to the symmetry of the lattice structure. It can also be seen that for the pure GCN, the dielectric function for the polarization parallel to $x$ axis is the same as that parallel to the $y$ direction, whereas for the B/P co-doped system , there is a slight difference. This difference can be attributed To the deformation occurred upon doping. Figure \ref{fig9}(e) illustrate the total optical absorption spectra for the pure and doped GCN monolayer. The pure GCN monolayer is found to have a strong absorption peak around 270-320 nm, attributed to the $\pi$-$\pi^*$ electronic transition, which can be commonly observed in the aromatic s-triazine compounds~\cite{srinivasu2014porous, li2006self}. Considering the inset of Fig. 9(a), the absorption edge of the undoped GCN is approximately 420 nm. In this regard, although the pure GCN is a visible light semiconductor photocatalyst, the visible light absorption is not sufficient to lead to the highly photocatalytic performance of this system. On the basis of optical absorption spectra of the doped systems, particularly P and B/P doped GCN monolayer indicate a very strong absorption tail (Urbach tail) in the visible region. A Similar behavior has been also reported recently~\cite{raziq2017synthesis, ran2015porous}. The reason behind the observed remarkable enhanced absorption in the visible region for P and B/P doped systems can be ascribed to the $\pi^*$ electronic transitions including lone pairs on the edge N atoms of the triazine rings. These transitions, prohibited in the planar pristine GCN, can be attributed to charge redistribution upon doping, caused by distorted configurable B/P co-doped GCN (Fig. \ref{fig9}(f)) (confirmed by both the electron density and Mulliken charge population). In other words, the forbidden $\pi^*$ electron transitions in the planar GCN are allowed in the distorted configuration mainly due to the deviation of the ring units from trigonal symmetry\cite{jorge2013h2, chen2014activation}.

The observed behavior by imaginay dielectric function is also confirmed by the optical conductivity measurement. This characteristic, which links the current density to the electric field, is one of the promising tools for studying the electronic states in materials. A redistribution of charges occurs when a system is subjected to an external electric field, resulting in current induction~\cite{lahiji2016first}. Figure \ref{fig10} shows the variation of optical conductivity as a function of the ultra-violate wavelength for pure, B doped, P doped, and B/P co-doped GCN monolayer. As it is exhibited in the figure, a rather sharp increase in optical conductivity takes place in the ultra-violate region and the correspondent peak moves toward lower wavelength with B, P and B/P doping. In the visible region, a kind of almost a linear trend of optical conductivity can be observed, which increases with B, P, and B/P doping. This shows the contribution of more electrons by the added dopants to the host material, resulting in higher optical conductance. According to the obtained results, it is believed that the B/P co-doped GCN monolayer can improve the utilization of visible portion of solar irradiation leading to high photocatalytic performance. This enhancement was experimentally observed by Raziq et al. for B and P co-doped GCN system~\cite{raziq2017synthesis}.

Aside from being a visible light active sample, the band edges of a suitable material for photocatalytic water splitting should be positioned appropriately with respect to the reduction and oxidation reaction levels of water in order to generate hydrogen and oxygen. The valence band maximum (VBM) and conduction band minimum (CBM) are calculated to be -3.40 and -6.10 eV, respectively. These values are in good agreement with experimental results~\cite{ma2016water}. The calculated CBM for the monolayer is similar to that of bulk GCN and it is found to be 1.12 eV above the water reduction level. However, the VBM shifts downward and is calculated to be 0.75 eV below the water oxidation level.
In the case of B/P co-doping, the VBM shifts remarkably upward, while the CBM shifts significantly downward as compared to the pure GCN monolayer. It should be noted that both pure and co-doped GCN nanosheet can be utilized in water splitting reactions. Although the pure system is more favorable for the reduction-oxidation reaction reactions, the co-doped GCN is expected to exhibit better photocatalytic performance due to the following reasons. (i) Using the visible  part of sunlight because of the lower band gap energy; (ii) increasing the absorption coefficient in the visible region because of activating $\pi^*$ electronic transitions in the distorted configuration; (iii) prolonging the lifetime of photo-excited electron-hole pairs owing to a partially or totally separated HOMO and LUMO.

 \section{Conclusion}\label{sec:concl}

We have studied and compared the electronic and optical properties of s-traiazine based graphitic carbon nitride (GCN) monolayer mono-doped with B and P as well co-doped with B/P using density functional theory calculations. Single layer GCN 2D system is found to have an increased band gap of 3.10 eV in comparison to that of 2.7 eV of the bulk GCN due to the quantum confinement effect. B-doped GCN monolayer exhibited a metallic character, while P-doped system showed a \textcolor{red}{metallic} behavior. Therefore, both systems were not suitable for using them in photocatalytic applications. Interestingly enough, the co-doped system displayed an appropriate band gap of 1.95 eV, making this configuration a promising candidate for water splitting reaction. Moreover, since more activating $\pi^*$ electronic transitions in the distorted configurations are observed, the optical absorption coefficient of the P-doped and B/P co-doped systems increased in the visible region which are beneficial in photocatalytic applications.

We remark that, in the accurate study of the optical properties of co-doped GCN, a model going beyond the DFT such as GW-DFT or time-dependent DFT simulations might be necessary to account for increasing correlation effects.

\section{Acknowledgment}
We would like to thank S. Tafreshi,  S. Yousefzadeh, and  M. Sabzali for useful discussions.  In addition, financial assistance of the Research and Technology Council of the Sharif University of Technology, support of the Iran National Science Foundation through Research Chair Award of Surface and Interface Physics Grant No. 940009 and the Iran Science Elites Federation Grant no 11/66332 are highly appreciated.

\bibliographystyle{aipnum4-1}
   \bibliography{manuscriptreview}
\end{document}